\documentclass[review]{elsarticle}

\usepackage{lineno,hyperref}
\usepackage{graphicx,epsf}
\usepackage{amsfonts}
\usepackage{amssymb}
\usepackage{amsmath}
\usepackage{overpic}
\usepackage{subcaption}
\usepackage{fancyhdr}
\modulolinenumbers[5]
\usepackage{framed}
\usepackage{lipsum,color}
\definecolor{shadecolor}{rgb}{1.,1.,1.}
\journal{Nucl.\,Phys.\,B}

\newenvironment{MyBox}{%
   \MakeFramed{\setlength{\hsize}{1.3\textwidth} \FrameRestore}}%
{\endMakeFramed}
\begin{document}

\begin{frontmatter}

\title{Topological phase transitions in four dimensions}

\author{Nicol\`o Defenu}
\address{Institute for Theoretical Physics, ETH Z$\ddot{u}$rich, Wolfgang-Pauli-Str. 27, 8093 Z$\ddot{u}$rich, Switzerland}
\author{Andrea Trombettoni}
\address{Department of Physics, University of Trieste, Strada Costiera 11, I-34151 Trieste, Italy}
\address{CNR-IOM DEMOCRITOS Simulation Center and SISSA, Via Bonomea 265, I-34136 Trieste, Italy}
\author{Dario Zappal\`a}
\address{INFN, Sezione di Catania, Via Santa Sofia 64, 95123 Catania, Italy}
\begin{abstract}
We show that four-dimensional systems may exhibit a topological phase
transition analogous to the well-known Berezinskii-Kosterlitz-Thouless
vortex unbinding transition in two-dimensional systems. 
We study a suitable generalization of the sine-Gordon model in four dimensions and the renormalization group flow equation of its couplings,
showing that the critical value of the frequency is
the square of the corresponding value in $2D$. 
The value of the anomalous dimension at the critical point is determined ($\eta=1/32$) and a conjecture for the universal jump
of the superfluid stiffness ($4/\pi^2$) presented.
\end{abstract}
\begin{keyword}
Topological phase transitions, High dimensions, Renormalisation Group.
\end{keyword}
\end{frontmatter}

\section{Introduction}

The introduction of an effective low-energy Hamiltonian for topological degrees of freedom, in order to describe their 
phase transition, is a conventional characteristic of  two dimensional systems. 
Yet, one may expect that a similar scenario may also occur in higher even dimension under specific conditions. 
This perspective leads to several open problems regarding the possible appearance of topological excitations in 
higher dimensions and, also, wether they may influence the physics in the four-dimensional relativistic space-time. 
More specifically,  a main question that one may pose are: \emph{"Are there topological phase transition in four dimensions?"} 
In the following we shall answer this question in an effective low-energy model, 
by showing the existence of a four dimensional lattice model whose continuous limit in Euclidean space-time displays 
a topological phase transition analogous to the two dimensional Berezinskii-Kosterlitz-Thouless (BKT) transition. 

The paradigmatic example of a topological phase transition occurring in absence
of spontaneous symmetry breaking and therefore not characterized by a local order parameter is the BKT transition. Its remarkable properties, 
such as low-temperature power law correlations,  can be understood in terms of the unbinding mechanism of low energy topological excitations. In general, the 
topologically relevant degrees of freedom for interacting two-dimensional systems with continuous $U(1)$ symmetry are vortices. Below the 
critical temperature, $T_{BKT}$,  vortices with opposite vorticity form pairs, that unbind above $T_{BKT}$. The mechanism of  vortex unbinding 
and the features peculiar of the BKT transition have been studied in a variety of different physical systems, ranging from $He$ films\,
\cite{Bishop77}, superconducting films\,\cite{Epstein81,Resnick81} and arrays of superconducting grains\,\cite{Martinoli00,Fazio01} to two-
dimensional systems of ultracold interacting bosons\,\cite{Hadzibabic06,Schweikhard07} and fermions\,\cite{Murthy15} as well as 1-dimensional topological quantum systems\,\cite{Sarkar2020}.


Indeed, the detection of the BKT transition can be done in different ways according to the specific system at hand: from the decay of the 
correlations functions, or from the observation of vortex unbinding, or from measurements of the superfluid fraction, or even from the 
scaling of the magnetization in finite size samples\,\cite{Bramwell94}.
Despite this wide range of observables, a remarkable common property,
specific of the BKT universality class, is the universal jump of the superfluid fraction (or, equivalently, the spin stiffness)
at the transitions temperature $T_{BKT}$. The amount of the jump is equal to $2/\pi$ \cite{Nelson77} and
related to the universal value of the critical exponent $\eta$ at the BKT
point: $\eta=1/4$. These properties can be studied
in the $2D$ spin model exhibiting BKT transition, the $XY$ model\,\cite{Kosterlitz1972,Kosterlitz1973,Jose1977}, as reviewed in\,\cite{Gulacsi98}.
Notice that the value of $\eta$ for temperatures between
$T=0$ (at which $\eta=1$) and the BKT critical temperature, $T=T_{BKT}$ 
(at  which $\eta=1/4$) is {\it not} universal,
and it depends on the specific model. A complete understanding of this
critical behaviour can be obtained by mapping in $2D$ the $XY$ model --
or more precisely, its low temperature limit, the Villain model,
which is in the same universality\,\cite{Jose1977,Villain75,Kleinert89} -- into the $2D$ Coulomb gas\,\cite{Minnhagen87},
which in turn can exactly be mapped onto the $2D$ sine-Gordon model\,\cite{Malard13}. The latter is a field theory
with an interaction term proportional to $\cos(\beta \phi)$ where $\beta$ is the frequency. The $1+1$
sine-Gordon model has been thoroughly investigated by several techniques,
including bosonization\,\cite{Gogolin04,Giamarchi2004}, functional renormalization group\,\cite{Nandori2001,Nagy2009,Nandori09,Daviet19}
and integrable approaches\,\cite{korepin_1993,bajnok_2013,Mussardo20}. The
main result is that there is a phase transition
occurring at a critical value of $\beta$, given by $\beta_c^2=8 \pi$
\cite{Coleman75}, which corresponds to the BKT superfluid transition.

These mappings are specific of two dimensions ($d=2$), and despite the sine-Gordon model and the Coulomb gas
can be mapped between them
in any dimension\,\cite{Kosterlitz77}, it is their mapping to the $XY$ model or to interacting bosons that is no longer valid in $d>2$. 
So, the properties of the BKT transition --
namely the presence of a line of fixed points, the
absence of magnetization, the presence of superfluidity in absence of condensation, and the universal jump
of the superfluid fraction -- are considered the hallmarks of phase transitions
in $2D$ systems.

In this paper we want to investigate how to obtain a BKT phase transition in
$d>2$, focusing on $d=4$. Despite BKT-like deconfinement properties in $d=3$\,\cite{Kleinert03} and some properties of the
isotropic Lifshitz points in $d=4$ have been considered and discussed \,\cite{Selke1978, Zappala18, Zappala19}, 
to the best of our knowledge
the remarkable features of the BKT universality class, such as the jump of the superfluid fraction and the
universality of the critical exponent $\eta$ at the end-point of the fixed points line have not been discussed
in $d>2$ or related to any realistic microscopic model. 
Here, we focus on $d = 4$ and determine in this case the universal properties of the BKT transition,
through the analysis of a sine-Gordon model that includes higher order derivative terms,
namely terms containing four spatial derivatives of the field.

The specific choice of higher derivative models in $d=4$ is prompted
by the long lasting analysis of the Lifshitz scaling and the related fixed point
structure\cite{Hornreich:1975zz}. In particular, the isotropic Lifshitz scaling arises when 
the standard two derivative term  in the action, $Z\,\partial \phi\partial \phi$, vanishes by setting 
$Z=0$, and the subsequent term $W\,\partial^2 \phi\partial^2 \phi$ becomes the leading 
derivative term, substantially modifying the standard dimensional scaling of the operators. 
As a consequence, a fixed point of the renormalization group for the $O(N)$ theory is expected  
when $4\,<\,d\,<\,8$  \cite{Hornreich:1975zz,Hornreich:pla}.
This expectation is supported by standard techniques such as the 
$1/N$ expansion \cite{Hornreich:pla,Shpot:2004gc,Gubser:2017vgc,Zappala:2017vjf}
or the $\epsilon$-expansion both below $d=8$ and above $d=4$ \cite{ PhysRevB.17.3607, Shpot:2001pk, Shpot:2004gc},
that confirm for the $O(N)$ theory
the role  of $d=8$ and $d=4$  as upper and lower critical dimensions, respectively. 
However, it must be noticed that the expansion around the 
lower critical dimension $d=4$ presents singularities  in $N=2$ \cite{Shpot:2004gc} , 
exactly as it happens for the standard expansion 
above $d=2$ \cite{Brezin:1975sq},
and therefore the case $N=2$  in $d=4$ requires alternative
approaches, likewise  the BKT transition in $d=2$.

In Sec.\,\ref{mic_model} we outline the microscopic lattice Hamiltonian, whose low energy theory shall display the aforementioned unbinding mechanism in $d=4$, and we briefly discuss a possible realisation in cold atom systems. Sec.\,\ref{f_th} is devoted to the study of the 4D sine-Gordon model, which we identified as a proper low energy theory to describe high-dimensional topological phase transitions. In order to further investigate the analogy with the vortex unbinding mechanism, in Sec.\,\ref{top_conf} the topological configurations driving the transition in $d=4$ are proposed and investigated in connection with the universal properties of the sine-Gordon model. Finally in Sec.\,\ref{concl} we discuss the possible applications of topological unbinding in 4D and we outline the future perspectives of the present investigations.
\section{The microscopic model}
\label{mic_model}

One of the most celebrated realization of BKT critical behaviour is the XY model on a square lattice. Here we will focus on its second neighbours generalization
\begin{equation}
H=-K \sum_{\langle{i,j}\rangle} \cos{(\theta_i-\theta_j)}- \tilde{K} \sum_{\langle\langle{i,j}\rangle\rangle}  \cos{(\theta_i-\theta_j)},
\label{H_mic} \end{equation} 
where $i,j$ denote the sites of a $4D$ lattice and $K=J/k_B T$, $\tilde{K}=\tilde{J}/k_B T$ with $J,\tilde{J}$
respectively the nearest-neighbour (n.n.) and next-nearest-neighbour (n.n.n.)
couplings. The partition function
is  $Z=\int \prod_i d\theta_i \, e^{-H}$. In the continuum limit, the action will contain both quadratic and quartic momentum contributions, 
due to the presence of n.n.n. couplings.
However, with the choice $\tilde{K}=-K/6$, ($K>0$), at mean field level one cancels  in (\ref{H_mic})
the quadratic momentum contributions,
so that the interacting $3+1$ field theory near to the critical point can be  described by\,\footnote{For a discussion of the effect of longer distance, including third neighbours couplings in cubic-lattice spin models see Ref.\,\cite{Selke1978} }:
\begin{align}
\label{4dsgbare}
\mathcal{S}[\varphi]=\int\left\{\frac{(\Delta\varphi)^{2}}{2}+g_{0}(1-\cos(\beta\varphi))\right\}d^{4}x
\end{align}
where $\Delta$ indicates the $4D$ Laplacian and $\varphi(x)$ is a real scalar field. 
The action (2), already considered in the context of $2D$ and $3D$ quantum dimer models\,\cite{Nogueira2009} 
and of models of $4D$ simplicial quantum gravity\,\cite{Mottola1997, Catterall1999}, will be studied in the following. 

We pause here to comment about possible connections with experimental setups and the requirements needed in principle 
  to have\,\eqref{H_mic}. 
  One could think to implement $3D$ quantum models at $T=0$ to emulate $4D$ classical systems at finite temperature \cite{Sachdev:404196} with the desired
  action as target.  
  So one at first sight could take
a $3D$ network of {\it quantum} Josephson junctions and add to them n.n.n.
interactions 
to emulate the model\,\eqref{H_mic} and therefore\,\eqref{4dsgbare}. A very clear discussion of this for $1+1$ quantum chains 
is done in \cite{Bradley84}, and reviewed in \cite{Sondhi97}. The result of this analysis is that one may have fourth derivatives
in the three spatial directions, but usual second derivative in the imaginary time direction. If from one side this is a case
interesting in itself, possibly in connection with tuning mechanisms of couplings along the imaginary time axis, from the other
side it clarifies that using quantum Josephson junctions with n.n.n. interactions appears not the best way
to realize\,\eqref{H_mic}, unless one does not come up with a proposal for the quantum emulation of higher order derivative
in the imaginary time direction. One may anyway resort to the proposal of implementing  lattices in synthetic dimensions
\cite{Boada12}, experimentally realized with cold $Yb$ atoms\,\cite{Mancini15}. 
In these schemes, the fourth direction could be realized by a large number of internal degrees of freedom, such as the $Yb$ levels. 
Remind that the Bose-Hubbard model can be mapped in the quantum phase model, and that
in a suitable range of parameters (in which interactions are not vanishing,
but negligible with respect to Josephson energy), one gets the $XY$ model
\cite{Anglin01,Trombettoni05}.
Therefore, in order to have\,\eqref{H_mic}, one needs a term of the form
$b^\dag b$ acting on n.n.n. sites, and this as well in the
extra, synthetic dimension.

\section{Field theory study}
\label{f_th}
The action in Eq.\,\eqref{4dsgbare} contains only a periodic local potential term in analogy with the usual sine-Gordon theory used to describe BKT
physics in low dimensions\,\cite{Giamarchi2004,Benfatto2013}. Within this
framework, the parameter $\beta$ is related to the phase stiffness
of the model, while the parameter $g_{0}$ describes the fugacity of
the topological excitations. It is worth noting that in $d=2$ a formal mapping is possible only at low temperatures between the traditional $O(2)$ model and the quadratic $2D$ sine-Gordon model \cite{Jose1977}. In the next section we are going to show how the theory in Eq.\,\eqref{4dsgbare} can be connected with the $4D$ quartic $O(2)$ via the introduction of certain singular phase configurations. 

In order to construct the RG study of the action in Eq.\,\eqref{4dsgbare}
we will employ the functional RG approach. This RG technique derives from
the possibility to write an exact RG equation for the effective
action\,\cite{Polchinski1984,Wegner1973,Wetterich1993}, which may then
be solved by projecting it on a restricted theory space parametrised by a proper ansatz\,\cite{Berges2002, Delamotte:2011jk}. This approach has successfully produced a comprehensive picture of the universal properties of $O(N)$ field theories as a function of the dimension $d$ and the symmetry index $N$\,\cite{Codello2015, Defenu:2017el} reproducing all the exactly known features of the phase diagram\,\cite{Defenu2015} also in presence of long-range interactions\,\cite{Defenu2015, Defenu2016, Defenu:2017dc}. 

The non-perturbative study of topological phase transitions within functional RG
requires to describe, at non-perturbative level, 
  the coupling between topological and spin-wave
  fluctuations starting from the microscopic variables of the model, see the discussions in \,\cite{Jakubczyk2014, defenu2017-1} and refs. therein.
A study of the inclusion of spin-wave fluctuations
  on top of the
  usual BKT RG flow equations has been presented in functional RG formalism in Ref.\,\cite{Krieg}. 
Since in the present case we will focus mainly on universal quantities we can employ an ansatz of the same form as the bare action in Eq.\,\eqref{4dsgbare}, which only accounts for the low-energy topological excitations responsible for the unbinding mechanism,
\begin{align}
\label{4dsgrun}
\Gamma_{k}[\varphi]=\int\left\{\frac{w_{k}}{2}(\Delta\varphi)^{2}+g_{k}(1-\cos\varphi)\right\}d^{4}x,
\end{align}
but with the bare coefficients substituted by scale dependent ones.
An ansatz analogous to the one in Eq.\,\eqref{4dsgrun}  has been proven
to reproduce all the qualitative features of the BKT transition, including
the universal jump of the superfluid stiffness\,\cite{Nagy2009,defenu2017-1}
and to yield consistent results for the computation of the
$c$-function\,\cite{Bacso2015}. More complicated ansatz were also shown to yield quantitative insight into the spectrum of the model\,\cite{Daviet19}.

By projecting the functional RG equation for the effective action on the
restricted theory space parametrised by the ansatz in Eq.\,\eqref{4dsgrun}
one obtains
\begin{align}
\partial_{t}V_{k}(\varphi)&=\int\frac{d^{d}q}{(2\pi)^{d}}G(q)\partial_{t}R_{k}(q),\label{potfl}\\
\partial_{t}w_{k}&=\lim_{p\to0}\int_{-\pi}^{\pi}\frac{d\varphi}{2\pi}\int\frac{d^{d}q}{(2\pi)^{d}}\partial_{t}R_{t}(q)G(q)^{2}V'''_{k}(\varphi)^{2}
\frac{d^{4}}{dp^{4}}G(p+q),\label{wfl}
\end{align}
where $t=-\log(k/\Lambda)$ is the RG logarithmic scale, $V_{k}(\varphi)=g_{k}(1-\cos\varphi)$ the local potential and $G(q)$ the propagator in momentum space
\begin{align}
G(q)=\frac{1}{w_{k}q^{4}+V''_{k}(\varphi)+R_{k}(q)}.
\end{align}
The function $R_{k}(q)$ is a regulator function which  introduces a finite
mass for long wave-length fluctuations $R_{k}(q\simeq 0)\approx k^{4}$.
The computation can be carried in $d$ dimensions, leading to the introduction of the generalized flow equations
\begin{align}
\partial_{t}w_{k}&=\beta w(w,g,d),\label{betaw}\\
\partial_{t}g_{k}&=\beta g(w,g,d).\label{betag}
\end{align}
In order to obtain an explicit form for the functions $\beta w$ and $\beta g$, it is convenient to introduce the regulator function $R_{k}(q)=k^{4}$, which allows to 
calculate the integrals in Eqs.\,\eqref{potfl} and\,\eqref{wfl} analytically. However, this choice for the regulator generates ultraviolet divergencies of the momentum 
integrals in $d=4$.  These divergencies are regularised by pursuing the computation for $d>4$ and, then taking the $d\to 4^{+}$ limit. The explicit calculation is 
shown in the\,\ref{fl_app}.

\begin{figure}[ht!]
\centering
\includegraphics[scale=.5]{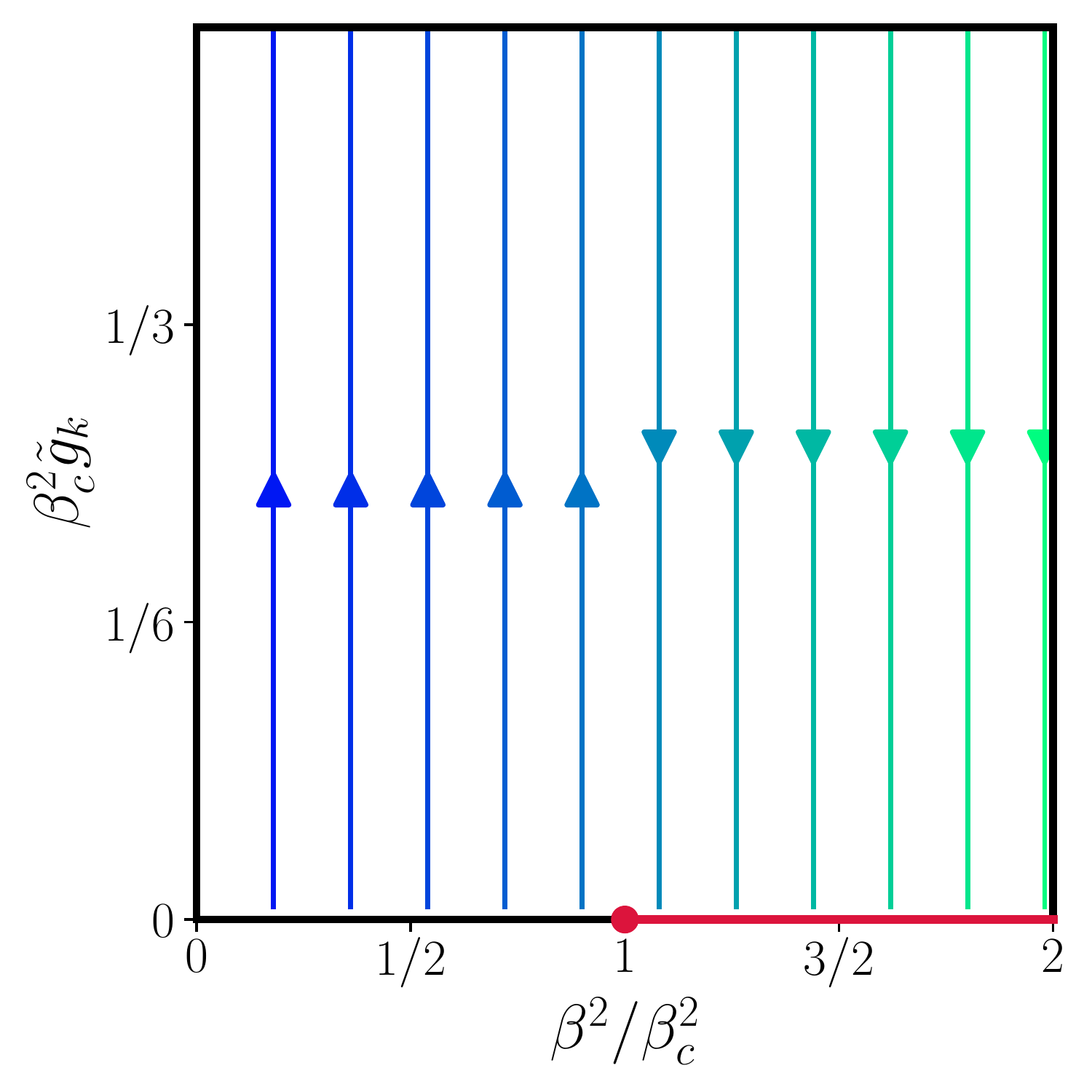}
\caption{The phase diagram at leading order
   in $g_k$ obtained by the
  evolution 
  Eq.\eqref{fl2} in the space of the dimensionless running
parameters $\tilde{g}_{k}$ and the sine-Gordon frequency $\beta^{2}$ both rescaled in terms of the critical 
frequency value $\beta_{c}^{2}=64\pi^{2}$. 
The similarities with the traditional BKT picture are evident: for $\beta^{2}>\beta_{c}^{2}$ one has a line of attractive fixed points with $g_{k}=0$, where the
system is massless. Conversely, if
$\beta^{2}<\beta_{c}^{2}$ the field theory
becomes massive and the flow is  attracted to a spinodal
 point at finite $\tilde{g}_{k}$.}
\label{Fig1}
\end{figure}

After introducing the rescaled variable $\tilde{g}_{k}=g_{k}/k^{4}$,
deferring the derivation to Appendix A, one finds
\begin{equation}
  (4-\partial_{t})\tilde{g}_{k}=\frac{1}{8\pi^{2}w_{k}\tilde{g}_{k}}\left(1-\sqrt{1-\tilde{g}_{k}^{2}}\right),\label{fl2}
  \end{equation}
 having a form similar to the $d=2$
   case\,\cite{Kosterlitz1972,Kosterlitz1973,Halperin1978}.
   At leading order in $\tilde{g}_{k}$ the running of the kinetic coupling $w_{k}$ vanishes and one can safely impose $w_{k}\equiv\beta^{-2} $ and employ the transformation $\tilde{g}_{k}\to\beta^{2}\tilde{g}_{k}$ in order to reduce the flow equations to the traditional form for the sine-Gordon model, see Ref.\,\cite{Bacso2015}. The resulting phase diagram displays a line of attractive Gaussian fixed points
 for $\beta^{2}>\beta_{c}^{2}$ with $g_{k}=0$, while for $\beta^{2}<\beta_{c}^{2}$ the $\cos(\varphi)$ perturbation becomes relevant and the 
 flow is driven at an infrared point with exponential correlations, see Fig.\,\ref{Fig1}.  Going beyond the leading order result in Eq.\,\eqref{fl2} one has to specify the flow equation for 
   $w_k$, which -- omitting algebraic details -- finally is found to be of
   the functional form $\partial_{t}w_{k}=C
   \frac{\tilde{g}_{k}^{2}}{(1-\tilde{g}_{k}^{2})^{\frac{3}{2}}}$ in formal analogy with the $d=2$ case.
 The value of the coefficient $C$ is the result of additional contributions not present in the  $d=2$ case, its calculation and value are reported in \ref{fl_app}. Notice that the sign of $C$ in $d=4$ found with the regulator $R_k=k^4$ is
   opposite to the corresponding one found in $d=2$ with the same kind of regulator (i.e., $R_k=k^2$). Further comments on this point require the
   analysis of the regulator-dependence of $C$.
 
 The critical value of the frequency in $d=4$, obtained from Eq.\,\eqref{fl2}, is
 \begin{equation}
   \beta_{c}^{2}=64\pi^{2}
   \label{quadrato}
 \end{equation}
in agreement with the heuristic arguments given in the next section. 
This value is universal and independent from the choice of the regulator, as it can be proven by expanding Eqs.\,\eqref{potfl} and\,\eqref{wfl} around $g_{k}=0$. 
Remarkably, the result\,\eqref{quadrato} is found to be the square of the corresponding standard result for the $2D$ sine-Gordon model, 
reading $\beta_c^2=8\pi$ \cite{Coleman75}.

The action in Eq.\,\eqref{4dsgbare} does not contain any quadratic momentum terms, as they vanish in the Hamiltonian 
in Eq.\,\eqref{H_mic} for $K=-6\tilde{K}$. 
Indeed, in order for the system to attain BKT behaviour, one has to tune two parameters: the temperature, which controls the $\beta$ parameter, 
and the nearest neighbour coupling $K$. 
Then, the BKT line of fixed points described by
Eq.\,\eqref{fl2}
is actually a line of third order critical points,
in analogy with the case of an isolated Lifshitz point\,\cite{Hornreich1980}. 
Yet, the actual critical value for the coupling $K>0$, may differ from the mean field value $K_{c}=-6\tilde{K}$ and, possibly, become temperature dependent. 
This specific critical value $K_{c}$ in the microscopic model described by Eq.\,\eqref{H_mic} is not a universal quantity and cannot be estimated by the 
continuum theory. Its determination by numerical simulations of the lattice Hamiltonian is left for future investigations.
In the following, we show  how the sine-Gordon theory
described here can be connected with the $4D$ quartic $U(1)$ model
by a suitable identification of the topological excitations. 

\section{Topological configurations}
\label{top_conf}
Now we illustrate the  example of a specific field configuration of a
$4D$ low energy  effective Hamiltonian for a $U(1)$ symmetric 
model with four derivatives of the field, that realizes the above picture. The effective  Hamiltonian is \begin{equation}
H[\vartheta({\bf r}) ]=\frac{\cal K}{2} \int d^4{\bf r}  \left [ \Delta \vartheta({\bf r})  \, \Delta  \vartheta({\bf r})  \right ] \; ,
\label{hamilU1} \end{equation} 
where $\mathcal{K}$ is the coupling and the field $\vartheta$ is the phase of a
complex scalar field $\Phi$, represented in polar components by $\vartheta$
and its  radial component ($\rho=\sqrt{\Phi\Phi^*}$). Fluctuations of $\rho$ are absent  in Eq.\,(\ref{hamilU1})
because they are suppressed in the infrared region by the presence of  a radial mass.
We notice that this suppression is warranted by the presence of a $\partial \Phi \partial \Phi^*$ term which,
in turn, yields a square momentum contribution in the propagator\,\cite{Popov1987};   however, in analogy with the 
criterion adopted for Eq.\,(\ref{4dsgbare}),  we discarded in Eq.\,(\ref{hamilU1}) 
the quadratic contribution $\partial \vartheta \partial \vartheta$, 
as this operator, if suitably taken on the critical manifold, is expected to be driven to zero by the RG flow in the low energy regime\,\cite{Zappala18}.
In addition, we did not include the term  $(\partial \vartheta \partial \vartheta)^2$, as it is possible to arrange the 
complex field four-derivative sector in such a way that only quadratic terms in $\vartheta$ are left.

We expect that the desired configuration  $\vartheta ({\bf r})= {\cal G} ({\bf r}-  {\bf r'} )$,
associated to a particular point ${\bf r'} $ in the $4D$  space, is such that 
$\Delta_{\bf r}  {\cal G} ({\bf r}-  {\bf r'} )= - ({\bf r} - {\bf r'})^{-2}$, 
as it  produces a logarithmic scaling of the  energy.
Then, from the solution of the Laplace equation  $-\Delta_{\bf r}  \, ({\bf r} - {\bf r'})^{-2} = (2\pi)^2 \delta^4 ({\bf r} - {\bf r'})$,  \,\cite{Evans},
we find 
\begin{equation}
{\cal G} ({\bf r}-  {\bf r'} ) =  \int \frac{d^4{\bf r''} }{(2\pi)^2}  \frac{1}{({\bf r} - {\bf r''})^{2}} \frac{1}{({\bf r''} - {\bf r'})^{2}} =
{\rm  ln}  \frac{R^{\frac{1}{2}} }{ |{\bf r} - {\bf r'}|^{\frac{1}{2}} } 
\label{config} 
\end{equation} 
where  $R$ is a large distance cutoff.
We remark that, starting with the Hamiltonian (\ref{hamilU1}) with  four field  derivatives, we are forced to work in $d=4$   to recover the 
logarithmic behavior of Eq. (\ref{config}), which is peculiar of this kind of transition; further details about  this issue are discussed 
in\,\ref{conf_app}.

Such scaling  is also realized by the  field configuration $\vartheta ({\bf r})=A_{\bf r'} ({\bf r})$
which has the following expression in terms  of spatial coordinates,
$A_{\bf r'} ({\bf r}) = (1/2)(\alpha_4 -\pi/2)\,{\rm cot}(\alpha_4)$, where $\alpha_4$ is the angle between  $({\bf r}-{\bf r'})$ and one of the 
the coordinate axes, e.g. ${\bf \hat x}_4$. We find that $A_{\bf r'} ({\bf r})$  is a solution of the equation  
$\Delta_{\bf r}  A_{\bf r'}  ({\bf r}) = -({\bf r} - {\bf r'})^{-2}$ and therefore, when inserted in Eq. (\ref{hamilU1}), it produces  equivalent  
effects to those of  ${\cal G} ({\bf r}-  {\bf r'} )$ (see\,\ref{conf_app}).

Consequently, we get $ \Delta_{\bf r}^2\, A_{\bf r'}  ({\bf r}) = (2\pi)^2 \delta^4 ({\bf r} - {\bf r'})$, i.e.
$A_{\bf r'}  ({\bf r})$, which is singular  at the point ${\bf r'}$, provides an extremum 
of the Hamiltonian (\ref{hamilU1}). The corresponding  energy is 
$H [ A_{\bf r'}  ] ={\cal K} \pi^2 \, {\rm  ln} \left ({R / r_0  } \right )$, $r_0$ being a short distance cutoff.
Then,  similarly to the 2-dimensional BKT transition, by estimating  the entropy as the logarithm of the number of ways to 
place $A_{\bf r'}$ (i.e. the point ${\bf r'}$) in the $d=4$ space with cut-offs $R$ and $r_0$:
$S[ A_{\bf r'}  ] ={\rm  ln} \left ({R^4 / r^4_0  } \right )$,  the free energy of the system exhibits  a change of sign  at 
\begin{equation}
{\cal K}_c=\frac{4}{\pi^2 } \; ,
\label{critical}
\end{equation}
to be associated with a measurable  discontinuous jump of ${\cal K}$ from $4/\pi^2$ to $0$.

In principle, by simple dimensional analysis, one can generalise the above argument to the case where the integrand in Eq. (\ref{hamilU1}) 
is replaced by  $\left [ \partial^m \vartheta \,\partial^m \vartheta  \right]^2$ and the integral is performed 
in a $2m$-dimensional space instead of $d=4$, with a suitably chosen function $ \vartheta({\bf r})$, such that
the Hamiltonian reduces to $ H^{(m)}={\cal K}^{(m)}/{2} \int d^{2m}{\bf r} \; ({ r})^{-2m} $ that is easily evaluated together 
with the associated entropy in the $2m$-dimensional space.
This straightforwardly leads to the determination of the critical  value ${\cal K}^{(m)}_c=2 (m!) / {\pi^m }$ which is the generalisation
of Eq. (\ref{critical}), obtained for $m=2$, and of the BKT transition ${\cal K}^{(1)}_c=2/\pi$, with $m=1$.
However, it must be remarked that this is the result of a simple dimensional analysis and a proper
generalisation of the resolution of the problem with $m > 2$ would require a thorough analysis
   including a full RG study as well as the explicit determination of relevant topological configurations in $d = 2m$.

Finally, by following a heuristic procedure already developed in   the  $2D$ case,
we can  map the sine-Gordon model in Eq.\,(\ref{4dsgrun}) 
onto Eq.\,(\ref{hamilU1}), computed for $\vartheta ({\bf r})= {\cal G} ({\bf r}-  {\bf r'} )$, 
and derive the relation $(2\pi)^4 {\cal K}=w^{-1}$ between the respective couplings. 
Details are displayed in\,\ref{rel_app}. Then, ${\cal K}_c$  in (\ref{critical}) corresponds to  
$\beta^2_c=64\pi^2$, in agreement with Eq.\,\eqref{quadrato}.

We are now able to determine the universal exponent  $\eta$, associated to the critical value ${\cal K}_c$. In fact, the
fixed point action in the low temperature phase (${\cal K}<{\cal K}_c$)
is simply Gaussian (see Fig. \ref{Fig1}),
and one can explicitly obtain the correlation functions of the vertex 
operator $V({\bf r})=\exp(i n\vartheta({\bf r}))$ \begin{align}
\langle V({\bf r})V(0)\rangle=\exp\left(-\frac{n^2}{2}  \langle \left( \vartheta({\bf r}) - \vartheta(0) \right)^2\rangle\right ).
\end{align}
From the correlation functions above one obtains the scaling of the vertex operator $\Delta_{n=1}=(8\pi^{2} {\cal K})^{-1}$, 
which can be compared with the conventional  $d=2$ result $\Delta^{2D}_{n=1}=(2\pi  {\cal K}_{2D})^{-1}$\,\cite{Fradkin2013}. 
As for the $2D$ case, the scaling of the vertex operator is connected with the power law decay of the correlation functions of the 
model, and this gives the following anomalous dimension associated to ${\cal K}_c$ :
\begin{align}
\eta=\frac{1}{32}
\end{align}
which has to be compared with the traditional BKT result $\eta_{2D}=1/4$.

\section{Conclusions}
\label{concl}
We showed that four-dimensional systems may exhibit a topological phase transition which extends to higher dimensions 
the celebrated Berezinskii-Kosterlitz-Thouless (BKT) transition.
A brief discussion of an experimental setup which may realise the effective action in Eq.\,\eqref{4dsgbare} is presented. 
We introduced  a suitable generalization of the sine-Gordon model in four dimensions
and we perfomed a renormalization group
flow equation of its couplings. The critical value of the sine-Gordon frequency
($\beta_c^2=64 \pi^2$) and the value of the anomalous dimension at the
critical point ($\eta = 1/32$) are determined.
A delicate point is to put in relation the $4D$ sine-Gordon model and a
suitable $O(2)$ model. In two dimensions this duality\,\cite{Jose1977} is at the heart of the
whole BKT theory, based on the identification of the vortex degrees of freedom
with Coulomb charges and on the exact mapping between the sine-Gordon model and the Coulomb gas. 
In the considered $4D$ case  we presented a discussion of the topological configurations and,
relying on this analysis, we presented a conjecture for the universal jump of the superfluid stiffness.

A stimulating input for future investigations 
comes from the strong analogy observed between the results obtained in our approach and those discussed in Ref.\,\cite{Mottola1997}.   
Also in light of this analogy, it would be as well very interesting to pursue the modelization of microscopic systems realizable
  in experimental setups, such as ultracold atoms in synthetic dimensions schemes or Josephson junction arrays at low temperature,
to devise possible experimental proposals for implementing\,\eqref{H_mic} or \,\eqref{4dsgbare} (or variants of them). 
In addition, motivated by the anisotropic Horava-Lifshitz approach to gravity \cite{Horava2009}, we mention that
it would be interesting to consider the anisotropic limit of our problem, with an action  that does not contain the
fourth derivative term in some  direction, and this will possibly allow an extension of the present theory to 
zero temperature anisotropic quantum systems in d = 3.  
\section*{Acknowledgements}
We thank T. Enss and I. Nandori for useful discussions.
This work is supported by the CNR / HAS (Italy-Hungary) project
"Strongly interacting systems in confined geometries'' and  by the Deutsche Forschungsgemeinschaft (DFG, German Research Foundation)  
under Germany’s Excellence Strategy “EXC-2181/1-
390900948” (the Heidelberg STRUCTURES Excellence
Cluster).
\appendix

\section{Derivation of the FRG flow equations}
\label{fl_app}
Let us rewrite our effective action ansatz, see Eq.\,(3) in the main text,
\begin{align}
\Gamma_{k}[\varphi]=\int\left\{\frac{w_{k}}{2}\Delta\varphi(x)\Delta\varphi(x)+g_{k}(1-\cos(\varphi))\right\}d^{d}x \,.
\end{align}
In the following we will derive the RG flow equations for the two couplings $g_{k}$ and $w_{k}$ within the FRG formalism.
The flow equations of the potential and the two point function for a single field action, obtained by the flow of the effective action\,\cite{Wetterich1993, Berges2002, Delamotte:2011jk}, read
\begin{align}
\partial_{t}V_{k}(\varphi)&=\frac{1}{2}\int\frac{\partial_{t}R_{k}(q)d^{d}q}{w_{k}q^{4}+V_{k}''(\varphi)+k^{4}},\label{vflow}\\
\partial_{t}w_{k}=&\frac{\mathcal{P}_{0}}{4!}\lim_{p\to0}\int{\frac{d^{d}q}{(2\pi)^{d}}\partial_{t}R_{t}(q)G(q)^{2}V'''_{k}(\varphi)^{2}\frac{d^{4}}{dp^{4}}G(p+q)}\label{w_flow}.
\end{align}
where $\mathcal{P}_{0}=\frac{1}{2\pi}\int_{-\pi}^{\pi} \cdots d\varphi$ is a projector on the field independent space and the regularized single field propagator reads\,\cite{Nagy2009}
\begin{align}
G(q)=\frac{1}{w_{k}q^{4}+V''_{k}(\varphi)+k^{4}}
\label{reg_prop}
\end{align}
which has been obtained by the introduction of the purely massive regulator 
\begin{align}
R_{k}(q)=k^{4}.
\end{align}
It is convenient to introduce the variable $y=|q+p|^{2}$ and rewrite the momentum derivative in Eq.\,\eqref{w_flow} according to the transformations
\begin{align}
\label{q_derivative}
\frac{1}{2}\frac{d^{2}}{dp^{2}}&=\frac{1}{2}\left(\frac{d^{2}y}{dp^{2}}\frac{d}{dy}+\left(\frac{dy}{dp}\right)^{2}\frac{d^{2}}{dy^{2}}\right),\\
\frac{1}{24}\frac{d^{4}}{dp^{4}}&=\frac{1}{24}\left(3\left(\frac{d^{2}y}{dp^{2}}\right)^{2}\frac{d^{2}}{dy^{2}}+6\left(\frac{dy}{dp}\right)^{2}\frac{d^{2}y}{dp^{2}}\frac{d^{3}}{dy^{3}}+\left(\frac{dy}{dp}\right)^{4}\frac{d^{4}}{dy^{4}}\right),\label{q_derivative2}
\end{align}
where the derivatives of the $y$ variable at $p=0$ read
\begin{align}
\label{y_derivatives}
\frac{dy}{dp}\Bigl{|}_{0}&=2q\cos\theta,\\
\frac{d^{2}y}{dp^{2}}\Bigl{|}_{0}&=2.
\end{align}
Inserting the expression in Eq.\,\eqref{q_derivative2} into Eq.\,\eqref{w_flow}, the $\beta$-function for the $w_{k}$ splits into three contributions
\begin{align}
\partial_{t}w_{k}= {\mathcal{P}_{0}}\, \left[  T_{1}+T_{2}+T_{3} \right]
\end{align}
with
\begin{align}
T_{1}&=\frac{d}{2}\,c_{d}\int q^{d-1}dq\partial_{t}R_{t}(q)G(q)^{2}V'''_{k}(\varphi)^{2}G^{(2)}(q),\label{t1}\\
T_{2}&=2c_{d}\int q^{d+1}dq\partial_{t}R_{t}(q)G(q)^{2}V'''_{k}(\varphi)^{2}G^{(3)}(q),\label{t2}\\
T_{3}&=2\frac{c_{d}}{d+2}\int q^{d+3}dq\partial_{t}R_{t}(q)G(q)^{2}V'''_{k}(\varphi)^{2}G^{(4)}(q).\label{t3}
\end{align}
where $c_{d}=(4\pi)^{-d/2}/\Gamma(1+d/2)$.
After obtaining the derivatives of the regularised propagator in Eq.\,\eqref{reg_prop} with respect to the $y$ variable and inserting them into Eq.\,\eqref{t1} one obtains
\begin{align}
T_{1}&=\frac{d}{2}\,c_{d}\int q^{d-1}dq\partial_{t}R_{t}(q)G(q)^{2}V'''_{k}(\varphi)^{2}\left(8w_{k}^{2}q^{4}G^{3}-2w_{k}G^{2}\right)=\nonumber\\
&=-2d\,c_{d}k^{4}\int q^{d-1}dqV'''_{k}(\varphi)^{2}\left(8w_{k}^{2}q^{4}G^{5}-2w_{k}G^{4}\right)=\nonumber\\
&=-2d\,c_{d}k^{4}V'''_{k}(\varphi)^{2}\left(8w_{k}^{2}\tilde{\text{I}}^{d+3}_{5}-2w_{k}\tilde{\text{I}}^{d-1}_{4}\right)
\end{align}
where
\begin{align}
\tilde{\text{I}}^{n}_{m}=\int\frac{q^{n}dq}{\left(w_{k}q^{4}+V''_{k}(\varphi)+k^{4}\right)^{m}}=\frac{\Gamma \left(\frac{n+1}{4}\right) \Gamma \left(m-\frac{n}{4}-\frac{1}{4}\right)}{4
   \Gamma (m)w_{k}^{\frac{n+1}{4}}}(V''_{k}(\varphi)
   +k^{4})^{\frac{n+1-4 m}{4} } 
\end{align}
The same procedure can be followed for the second term
\begin{align}
T_{2}&=2c_{d}\int q^{d+1}dq\partial_{t}R_{t}(q)G(q)^{2}V'''_{k}(\varphi)^{2}G^{(3)}(q)=\nonumber\\
&=2c_{d}\int q^{d+1}dq\partial_{t}R_{t}(q)G(q)^{2}V'''_{k}(\varphi)^{2}\left(24w_{k}^{2}q^{2}G^{3}-48w_{k}^{3}q^{6}G^{4}\right)=\nonumber\\
&=-8c_{d}k^{4}V'''_{k}(\varphi)^{2}\left(24w_{k}^{2}\tilde{\text{I}}^{d+3}_{5}-48w_{k}^{3}\tilde{\text{I}}^{d+7}_{6}\right),
\end{align}
and the third term
\begin{align}
T_{3}&=2\frac{c_{d}}{d+2}\int q^{d+3}dq\partial_{t}R_{t}(q)G(q)^{2}V'''_{k}(\varphi)^{2}G^{(4)}(q)=\nonumber\\
&=-8\frac{c_{d}}{d+2}k^{4}\int q^{d+3}dq\,G(q)^{2}V'''_{k}(\varphi)^{2}(384w_{k}^{4}q^{8}G^{5}-288w_{k}^{3}q^{4}G^{4}+24w_{k}^{2}G^{3})=\nonumber\\
&=-8\frac{c_{d}}{d+2}k^{4}V'''_{k}(\varphi)^{2}(384w_{k}^{4}\tilde{\text{I}}^{d+11}_{7}-288w_{k}^{3}I^{d+7}_{6}+24w_{k}^{2}\tilde{\text{I}}^{d+3}_{5}).
\end{align}
Let us define the two flow equations as follows
\begin{align}
\partial_{t}w_{k}&=\beta w(w,g,d),\label{betaw_sm}\\
\partial_{t}g_{k}&=\beta g(w,g,d).
\label{betag_sm}\end{align}
In order to pursue the computation of $\beta w$ we have to insert the parametrisation $V_{k}(\varphi)=g_{k}(1-\cos(\varphi))$ 
and take the integral in $\varphi$ from $-\pi$ to $\pi$. All terms have the same form and can be computed by defining the new quantities 
\begin{MyBox}
\begin{align}
\text{I}^{n}_{m}&=\int_{-\pi}^{\pi}\frac{V'''_{k}(\varphi)^{2}}{2\pi}\tilde{\text{I}}^{n}_{m}=\frac{g_{k}^{2}}{2\pi}\frac{\Gamma 
\left(\frac{n+1}{4}\right) \Gamma \left(m-\frac{n}{4}-\frac{1}{4}\right)}{4
 \Gamma (m)w_{k}^{\frac{n+1}{4}}}\int_{-\pi}^{\pi}\frac{\sin^{2}\theta}{\left(k^{4}+g_{k}\cos\theta\right)^{\frac{4m-n-1}{4}}}
=\frac{\Gamma \left(\frac{n+1}{4}\right) \Gamma \left(m-\frac{n}{4}-\frac{1}{4}\right)}{
   \Gamma (m)w_{k}^{\frac{n+1}{4}}}\nonumber\\
&\cdot \frac{2 \left(k^8-k^4
   g_{k}\right) \, _2F_1\left(-\frac{1}{2},m-\frac{n}{4}-\frac{1}{4};1;-\frac{2
   g}{k^4-g}\right)-\left(k^4+g_{k}\right) \left(2 k^4+g_{k} (4 m-n-5)\right) \,
   _2F_1\left(\frac{1}{2},m-\frac{n}{4}-\frac{1}{4};1;-\frac{2
   g_{k}}{k^4-g_{k}}\right)}{ (4 m-n-9) (4 m-n-5)\left(k^4-g_{k}\right)^{\frac{4 m-n-1}{4}} }.
\end{align}
\end{MyBox}
Irrespectively of the choice of $m$ and $n$ we can define the rescaled parameter $\tilde{g}_{k}=g_{k}/k^{4}$ leading to
\begin{MyBox}
\begin{align}
\text{I}^{n}_{m}&=\frac{\Gamma \left(\frac{n+1}{4}\right) \Gamma \left(m-\frac{n}{4}-\frac{1}{4}\right)}{
\Gamma (m)w_{k}^{\frac{n+1}{4}}}k^{5+n-4m}\nonumber\\
&\cdot\frac{2 \left(1- \tilde{g}_{k}\right) \, _2F_1\left(-\frac{1}{2},m-\frac{n}{4}-\frac{1}{4};1;-\frac{2
\tilde{g}_{k}}{1-\tilde{g}_{k}}\right)-\left(1+\tilde{g}_{k}\right) \left(2 +\tilde{g}_{k} (4 m-n-5)\right) \,
_2F_1\left(\frac{1}{2},m-\frac{n}{4}-\frac{1}{4};1;-\frac{2
\tilde{g}_{k}}{1-\tilde{g}_{k}}\right)}{ (4 m-n-9) (4 m-n-5)\left(1-\tilde{g}_{k}\right)^{\frac{4 m-n-1}{4}} }
\end{align}
\end{MyBox}
A close inspection of the expressions for the terms $T_{1,2,3}$ reveals that they are all independent on $w_{k}$, so that the $\beta$-function for the $w_{k}$ parameter only depends on $g_{k}$. Then, expanding around $d=4$ one obtains the following expressions
\begin{align}
T_{1}&=-\frac{d-4}{192\pi^{2}}\frac{\tilde{g}_{k}^{2}}{(1-\tilde{g}_{k}^{2})^{\frac{3}{2}}}+O(d-4)^{2}\\
T_{2}&=-\frac{1}{80\pi^{2}}\frac{\tilde{g}_{k}^{2}}{(1-\tilde{g}_{k}^{2})^{\frac{3}{2}}}+O(d-4)\\
T_{3}&=-T_{2}/2.
\end{align}
In the traditional functional RG study of the sine-Gordon model in $d=2$ the $\beta$-function for $w_{k}$ would follow from a contribution of the type of $T_{1}$. However, such contribution vanishes in the present $d=4$ case and the entire contribution to $\beta w$ come from the higher order terms $T_{2}$ and $T_{3}$, which yield the same functional form as in $d=2$, but a coefficient with opposite sign.

Conversely, the flow of the coupling $g_{k}$ can be derived in full analogy with the $d=2$ calculation. Indeed, if  the potential is parametrized as
 $V_{k}(\varphi)=g_{k}(1-\cos(\varphi))$, then $g_{k}$ is obtained with the help of the following projector $\mathcal{P}_{1}$, as 
$g_{k}= \mathcal{P}_{1}[V_{k}(\varphi)]=\frac{-1}{\pi}\int_{-\pi}^{\pi}\,[V_{k}(\varphi)]\cos(\varphi) d\varphi$, and therefore
$\beta g$ is derived by applying  $\mathcal{P}_{1}$ to both sides of  Eq.\,(\ref{vflow}) :
\begin{MyBox}
\begin{align}
\beta g&=-\frac{1}{\pi}\int_{-\pi}^{\pi}\partial_{t}V_{k}(\varphi)\cos(\varphi)=
\frac{4k^{4}d\,c_{d}}{\pi}\int_{-\pi}^{\pi}\cos\varphi d\varphi\int_{0}^{+\infty}\frac{q^{d-1}dq}{w_{k}q^{4}+g_{k}\cos(\varphi)+k^{4}}=
\frac{4k^{4}d\,c_{d}}{\pi}\int_{-\pi}^{\pi}\tilde{\text{I}}^{d-1}_{1}\cos\varphi d\varphi\nonumber\\
&=-\frac{k^{4}}{w_{k}\sin\left(\frac{d\pi}{4}\right)}\int_{-\pi}^{\pi}\left(\frac{w}{k^{4}+g_{k}\cos(\varphi)}\right)^{1-\frac{d}{4}}\cos(\varphi)d\varphi=
\frac{k^{4}}{(g_{k}^{2}-k^{8})\sin\left(\frac{d\pi}{4}\right)}\left(\frac{k^{2}-g_{k}}{w_{k}}\right)^{d/4}\nonumber\\
&\cdot\left(3 \left(k^4-g_{k}\right) \, _2F_1\left(-\frac{1}{2},1-\frac{d}{4};2;-
\frac{2  g_{k}}{k^4-g_{k}}\right)-\left (3 k^{4}-(d-1)g_{k}\right ) \,
   _2F_1\left(\frac{1}{2},1-\frac{d}{4};2;-\frac{2 g_{k}}{k^4-g_{k}}\right)\right).\end{align}
   \end{MyBox}
Our focus is the description of the BKT scaling, which appears in case of marginal scaling of the couplings, then 
we take the $d\to4^{+}$ limit of the general $\beta$-functions, yielding
 \begin{align}
 \partial_{t}w_{k}&=\lim_{d\to4}\beta w(w,g,d)=-\frac{k^{4}}{160\pi^{2}}\frac{g_{k}^{2}}{(k^{8}-g_{k}^{2})^{\frac{3}{2}}},\\
  \partial_{t}g_{k}&=\lim_{d\to4}\beta g(w,g,d)=-\frac{k^{4}}{8\pi^{2}w_{k}g_{k}}\left(k^{4}-\sqrt{k^{8}-g_{k}^{2}}\right).
 \end{align}
 According to the transformation $g_{k}=k^{4}\tilde{g}_{k}$, the dimensionless flow equations read
 \begin{align}
   \partial_{t}w_{k}&=
   C \, \frac{\tilde{g}_{k}^{2}}{(1-\tilde{g}_{k}^{2})^{\frac{3}{2}}},\label{fl1_sm}\\
  (4-\partial_{t})\tilde{g}_{k}&=\frac{1}{8\pi^{2}w_{k}\tilde{g}_{k}}\left(1-\sqrt{1-\tilde{g}_{k}^{2}}\right)\label{fl2_sm}.
 \end{align}
 as reported in the main text (with $C=-\frac{\pi^{2}}{160}$)
\section{Topological configurations in d=4}
\label{conf_app}
In $d=4$, the coordinates in spherical representation are given by
\begin{eqnarray}
x_1&=&{\rm r}\,{\rm sin}(\phi_4)\,{\rm sin}(\phi_3)\,{\rm sin}(\phi_2)\nonumber\\
x_2&=&{\rm r}\,{\rm sin}(\phi_4)\,{\rm sin}(\phi_3)\,{\rm cos}(\phi_2)\nonumber\\
x_3&=&{\rm r}\,{\rm sin}(\phi_4)\,{\rm cos}(\phi_3)\nonumber\\
x_4&=&{\rm r}\,{\rm cos}(\phi_4)
\label{coordinates}
\end{eqnarray}
where ${\rm r}=\sqrt{
x_i\,x_i}$ and the angles $\phi_4, \; \phi_3$ are defined in the range $[0,\pi]$ while $\phi_2$ in the range $[0, 2\pi)$
and, by inverting the last line in Eq.\,(\ref{coordinates}), $\phi_4$ is 
\begin{equation}
\phi_4 ={\rm ArcCos}\left(\frac{x_4}{\rm r}\right) \;.
\label{fi4}
\end{equation}
Within this representation, the laplacian $\Delta=\partial_i \partial_i$  has the following expression
\begin{equation}
\Delta= \frac{\partial_{\rm r} [{\rm r}^3\partial_{\rm r}]}{{\rm r}^3} + \frac{\partial^2_{\phi_2}}{r^2\,{\rm sin}^2(\phi_4)\,{\rm sin}^2(\phi_3)}+
\frac{\partial_{\phi_3} [{\rm sin}(\phi_3) \partial_{\phi_3}]  }{{\rm r}^2\,{\rm sin}^2(\phi_4)\,{\rm sin}(\phi_3)} +
\frac{\partial_{\phi_4} [ {\rm sin}^2(\phi_4) \partial_{\phi_4}]  }{{\rm r}^2\,{\rm sin}^2(\phi_4)} \;.
\label{laplsc}
\end{equation}
Therefore, it is easy to verify that the scalar configuration $A ({\bf r})$,
\begin{equation}
A ({\bf r}) =\frac{1}{2} \left(\phi_4-\frac{\pi}{2}\right)\,\frac{{\rm cos}(\phi_4)}{{\rm sin}(\phi_4)}
\label{solorig}
\end{equation}
which is essentially related to the angle  $\phi_4$  between  ${\bf r}$ and  the coordinate axis  ${\bf \hat x}_4$, yields
\begin{equation}
\Delta A ({\bf r}) =- \,\frac{1}{{\rm r}^2}  \;.
\label{sollap}
\end{equation}
The configuration $A$ depends on  $\phi_4$ only, it is defined for $0<\phi_4< \pi$, and
the shift by $-\pi/2$ in the definition makes it symmetric in  the interval $[0,\pi]$ with respect to the point $\pi/2$.
The concavity of  $A$ turns  downward and $A<0$ everywhere, except at its maximum in $\phi_4=\pi/2$, where $A=0$.
The factor ${\rm cos}(\phi_4)/(2{\rm sin}(\phi_4))$ makes $A$ divergent to $-\infty$, both in $\phi_4=0$
and $\phi_4=\pi$, but it is essential to recover the spherical symmetry of   $\Delta A ({\bf r})$  shown in Eq.\,(\ref{sollap}).
In fact, when dealing  with  an Hamiltonian that contains the laplacian of the (real) field $\vartheta$ only:
\begin{equation}
H[\vartheta({\bf r}) ]=\frac{\cal K}{2} \int d^4{\bf r}  \left [ \Delta \vartheta({\bf r})  \, \Delta  \vartheta({\bf r})  \right ] \; ,
\label{hamil} 
\end{equation} 
the configuration $\vartheta({\bf r})=A ({\bf r})$  does not induce any singularity 
along the axis ${\bf \hat x}_4$ in the integrand in Eq.\,(\ref{hamil}), with  the exception of the point ${\bf r}=0$.
In addition, $A ({\bf r})$  corresponds to an extremal field configuration, as it
is verified  with the help of  Eq.\,(\ref{sollap})
and by recalling the solution of the Laplace equation in $d=4$\,\cite{Evans}:
\begin{equation}
\Delta  \, \frac{-1}{({\bf r} - {\bf r'})^{2}} = (2\pi)^2 \delta^4 ({\bf r} - {\bf r'}) \;,
\label{lapsol}
\end{equation}
which imply
\begin{equation}
\Delta^2\, A({\bf r}) =(2\pi)^2 \delta^4 ({\bf r})
\label{lapsquare}
\end{equation}
i.e. $\Delta^2\, A$  vanishes everywhere, with the exception of the point ${\bf r}=0$.

Therefore, due to  Eq.\,(\ref{lapsquare}),  the solution in Eq.\,(\ref{solorig}) can be regarded as the potential generated  by a charge located at the origin,
but with the standard laplacian replaced by the square laplacian. The integration of the left hand side of  Eq.\,(\ref{lapsquare}),
extended to any volume containing the origin ${\bf r}=0$,  gives $(2\pi)^2$, while it vanishes if ${\rm r}=0$ is external.
Obviously, one can introduce  a general configuration without modifying the results of the above analysis,
with the singularity  of Eq.\,(\ref{solorig}) in ${\bf r}=0$, shifted  to the generic point ${\bf r'}$,
\begin{equation}
A_{\bf r'} ({\bf r}) =\frac{1}{2} \left(\alpha_4-\frac{\pi}{2}\right)\,\frac{{\rm cos}(\alpha_4)}{{\rm sin}(\alpha_4)}
\label{solgener}
\end{equation}
and now $\alpha_4$ indicates the angle between  $({\bf r}-{\bf r'})$ and ${\bf \hat x}_4$.

The scaling displayed in Eq.\,(\ref{sollap}) by the configuration $A({\bf r}) $,
is also observed for 
\begin{equation}
{\cal G} ({\bf r}-  {\bf r'} ) =  \int \frac{d^4{\bf r''} }{(2\pi)^2}  \frac{1}{({\bf r} - {\bf r''})^{2}} \frac{1}{({\bf r''} - {\bf r'})^{2}} 
\label{configapp} 
\end{equation} 
where ${\bf r'}$ indicates the location of the singularity. In fact, from  Eq.\,(\ref{lapsol}) one finds 
\begin{equation}
\Delta_{\bf r}   \, {\cal G} ({\bf r}-  {\bf r'} ) =
\frac{-1}{({\bf r} - {\bf r'})^{2}} 
\label{lapsol2}
\end{equation}
and
\begin{equation}
\Delta_{\bf r}^2\, {\cal G} ({\bf r}-  {\bf r'} )  = (2\pi)^2 \delta^4 ({\bf r} - {\bf r'}) \;,
\label{lapsquare2}
\end{equation}
as observed for $A({\bf r})$ in Eqs.\,(\ref{sollap})   and (\ref{lapsquare}). This indicates that 
${\cal G} ({\bf r} - {\bf r'} )$  and  $A_{\bf r'} ({\bf r})$ can be interchanged in the  Hamiltonian in Eq.\,(\ref{hamil}) 
with no consequence.

In addition, by introducing a large distance spatial cut-off $R$,
the integral in Eq.\,(\ref{lapsol2}) can be solved :
\begin{equation}
{\cal G} ({\bf r} -  {\bf r'} ) = \frac { 1}{4}\; {\rm  ln} \frac{R^2}{ ({\bf r} - {\bf r'})^2 } \;,
\label{gvalue} 
\end{equation} 
indicating  that ${\cal G} ({\bf r} -  {\bf r'} )$ decreases from large positive values to zero, when the distance $|{\bf r} -  {\bf r'} | $
grows up to the cut-off $R$.  Clearly when the limit ${\bf r} \to  {\bf r'} $ is taken, the logarithm diverges and one must require 
the validity of the expression in (\ref{gvalue}) only up to a minimum distance ${\rm r}_0$ from the singularity in ${\bf r'}$.
Then,  from Eqs.\,(\ref{lapsquare2}) and (\ref{gvalue}), it is easy to calculate the energy of a single configuration 
(where, again, we make use of the ultraviolet cutoff ${\rm r}_0$)
\begin{equation}
H[{\cal G}]=  \frac{\cal K}{2} \pi^2 \; {\rm  ln} \frac{R^2}{{\rm r}_0^2 } \; .
\label{ener} 
\end{equation} 
The same result is obtained by directly computing $H$ with the help of Eq.\,(\ref{lapsol2}).

After computing the energy associated to a single charge, we consider the configuration associated to a distribution of charges located at different points
\begin{equation}
{\cal G}^C ({\bf r}) = \sum_i  \, n_i  \, {\cal G} ({\bf r} -  {\bf r_i} )  \;,
\label{distri} 
\end{equation} 
where $n_i \in Z$ indicates the number of (positive or negative) charges at the point ${\bf r_i}$ and the Hamiltonian is 
\begin{MyBox}
\begin{equation}
H[ {\cal G}^C]=\frac{\cal K}{2} \int d^4{\bf r}  \, \sum_{i,j}   \, n_i  n_j   \, \left [ \Delta {\cal G} ({\bf r} -  {\bf r_i} ) ][ \Delta{\cal G} ({\bf r} -  {\bf r_j} ) \right ]  =
\sum_{i} \, n^2_i  \epsilon_{_S} \, + \,
\frac{\cal K}{2} \int d^4{\bf r}  \, \sum_{i \neq j}   \, n_i  n_j   \, \left [ \Delta {\cal G} ({\bf r} -  {\bf r_i} )][\Delta{\cal G} ({\bf r} -  {\bf r_j} ) \right ] 
\label{hamildist} 
\end{equation} 
\end{MyBox}
where we isolated the contribution due to the self-energy, indicated with $\epsilon_{_S}$, related to the cases in which $i=j$ in the sum.
Then, it is straightforward to compute the second term in the right hand side of  Eq.\,(\ref{hamildist}), for the elementary case of two distinct charges,
one located in ${\bf r_a}$ and the other in ${\bf r_b}$ ($n_{a,b}=\pm 1$),  with the help of Eqs.\,(\ref{lapsquare2}) and (\ref{gvalue}),
\begin{eqnarray}
H_{a,b}&=&(n^2_a +n^2_b) \epsilon_{_S} + \frac{\cal K}{2} \int d^4{\bf r} \; 2 n_a  n_b   \, \left 
[ \Delta {\cal G} ({\bf r} -  {\bf r_a} ) ][ \Delta{\cal G} ({\bf r} -  {\bf r_b} ) \right ]  =
\nonumber\\
&&(n^2_a +n^2_b) \epsilon_{_S} +\frac{\cal K}{2} \int d^4{\bf r}  \; 2 n_a  n_b  \, (2\pi)^2 \delta^4({\bf r} -  {\bf r_a} )  \;  \frac { 1}{4}\; {\rm  ln} \frac{R^2}{ ({\bf r} 
- {\bf r_b})^2 }=
\nonumber\\
&&(n^2_a +n^2_b) \epsilon_{_S} - \frac{\cal K}{2}\, n_a  n_b  \, (2\pi)^2  \; {\rm  ln} \frac{ |{\bf r_a} - {\bf r_b}| }{R}
\label{hamil2ch} 
\end{eqnarray} 
Apart from the self-energy contribution the energy coming from the interaction of two distinct charges is positive (negative) if the product  $n_a  n_b $ is positive
(negative), according to the expectations,  because the distance between the two charges $|{\bf r_a} - {\bf r_b}|$ is smaller than the large distance cut-off $R$.

It must be noticed that the presence of the four derivatives in Eq.\,(\ref{hamil}), implies that the logarithmic scaling is peculiar of a  $d=4$ space. In fact,
if we repeat the above considerations in $d=3$, by replacing the Green function of the laplacian in Eq.\,(\ref{lapsol}) with the three-dimensional  
$-1/{|{\bf r} - {\bf r'}|} $, we find that the corresponding function ${\cal G}$ grows linearly with $R$ (instead of the  logarithmic growth of Eq.\,(\ref{gvalue})), 
as it can be checked by simple dimensional analysis. Moreover, in the $d=3$ case,
the linear dependence on $R$ is also found in the computation of the energy of the
configuration ${\cal G}$, instead of the logarithmic dependence found in Eq.\,(\ref{ener})).

\section{Relation with the sine-Gordon model}
\label{rel_app}

We now establish a relation between the Hamiltonian in Eq.\,(\ref{hamildist}) and the sine-Gordon model in $d=4$. For the moment we 
neglect the self-energy part, $\sum_{i} \, n^2_i  \epsilon_{_S} $ and focus on the remaining part that, according to the result in Eq.\,(\ref{hamil2ch}),
can be written as
\begin{equation}
H_{_I}  = - \frac{\cal K}{2} \, \sum_{i \neq j}   \, n_i  n_j \, (2\pi)^2 \frac { 1}{4}\; {\rm  ln} \frac{ ({\bf r_i} - {\bf r_j})^2 }  {R^2} =
\frac{\cal K}{2} \, (2\pi)^2  \int  d^4{\bf r}_1   \int  d^4{\bf r}_2   \; n({\bf r}_1) \;{\cal G} ({\bf r}_1 -  {\bf r}_2 )  \;   n({\bf r}_2)  \;,
\label{hr} 
\end{equation} 
where we introduced the charge density real field 
\begin{equation}
n({\bf r}) = \sum_{i } \, n_i \, \delta^4 ({\bf r} - {\bf r_i}) \; .
\label{enne} 
\end{equation} 
By introducing the Fourier Transform (FT) of ${\cal G} ({\bf r}_1 -  {\bf r}_2 )$ according to 
\begin{equation}
\widetilde{\cal G} ({\bf p }) =  \int  \frac{d^4{\bf r}}
{(2\pi)^2}  \,  e^{i\, {\bf p\cdot r }}\; {\cal G} ({\bf r}) = \left (\frac{1}{{\bf p}^2 }\right)^2\;,
\label{FT} 
\end{equation} 
and also the FT $\widetilde n ({\bf p })$ of the density field,
the Hamiltonian in Eq.\,(\ref{hr}) becomes
\begin{equation}
H_{_I}  = 
\frac{\cal K}{2} \, (2\pi)^4  \int  d^4{\bf p} \; \widetilde n ({\bf p }) \;  \widetilde{\cal G} ({\bf p }) \; \widetilde n (- {\bf p }) \; = 
\frac{1}{2}\int  d^4{\bf p} \; \widetilde n ({\bf p }) \; 
\frac{(2\pi)^4  {\cal K}}{ p^4}\; \widetilde n (- {\bf p })\;.
\label{hrFT}  
\end{equation} 
The partition function associated to $H_{_I}$ can be written by introducing 
a gaussian functional integration over an auxiliary field  $\widetilde \phi({\bf p })$, suitably inserted in the exponent (${\cal N} $ is the normalization factor) :
\begin{equation}
{\rm exp} \left [ - H_{_I} \right ] = {\cal N} \int D \widetilde \phi \;  {\rm exp} \left [ - \frac{1}{2} \int d^4{\bf p} \;
\frac{p^4 \, |\widetilde \phi({\bf p })|^2}{ (2\pi)^4  {\cal K} }
+ \frac{i}{2} \int d^4{\bf p} \left (   \widetilde \phi({\bf p }) \,  \widetilde n (- {\bf p })     +      \widetilde \phi(- {\bf p }) \, \widetilde n ({\bf p }) \right )
\right ]
\label{partitionFT} 
\end{equation} 
where it is understood that the inverse temperature factor $T^{-1}$ in the partition function is absorbed  here by the  redefinition  $ T^{-1} {\cal K} \to{\cal K}$.
Then, by taking the inverse FT, one finds
\begin{equation}
{\rm exp} \left [ - H_{_I} \right ] = {\cal N} \int D \phi \;  {\rm exp} \left [ - \frac{1}{2 \, (2\pi)^4  {\cal K} } \int d^4{\bf r} \;
\Delta \phi({\bf r})\, \Delta \phi({\bf r})
+ i \, \int d^4{\bf r} \left (  \phi({\bf r }) \,  n ({\bf r }) \right )
\right ]
\label{partition} 
\end{equation} 
The full partition function includes the additional term of the Hamiltonian neglected so far 
\begin{equation}
{\cal Z} = {\rm exp} \left [ - H_{_I}  - \sum_{i} \, n^2_i  \, \epsilon_{_S} \right ] 
\label{zeta}
\end{equation}
and we signal the dependence of the latter and of the second integral in the exponent in Eq.\,(\ref{partition}) on $n_i$, while the first integral 
does not depend on the number of charges. 

Then, one can sum in the partition function over all possible configurations with either zero charge or
one positive or one negative charge, located in a generic point ${\bf r}_s$, and discard the other configurations with two or more charges, that are 
negligible because exponentially suppressed :
\begin{MyBox}
\begin{eqnarray}
{\cal Z} &=& {\cal N} \int D \phi \;  {\rm exp} \left [ - \frac{1}{2 \, (2\pi)^4  {\cal K}} \int d^4{\bf r} \;
\Delta \phi({\bf r})\, \Delta \phi({\bf r}) \right ]   \; \left\{ 1   + \int d^4 {\bf r}_s  \;  {\rm e}^{ -  \epsilon_{_S} + i \phi({\bf r}_s ) }
+ \int d^4 {\bf r}_s  \;  {\rm e}^{ -  \epsilon_{_S} - i \phi({\bf r}_s )}   \right\}
\nonumber\\
&=& {\cal N} \int D \phi \;  {\rm exp} \left [ - \frac{1}{2 \, (2\pi)^4  {\cal K} } \int d^4{\bf r} \;
\Delta \phi({\bf r})\, \Delta \phi({\bf r}) \right ]   \; \left\{ 1   + \int d^4 {\bf r}_s  \; {\rm e}^{ - \epsilon_{_S} } \; 2 \;{\rm cos}( \phi({\bf r}_s ) )   \right\}
\nonumber\\
&=& {\cal N} \int D \phi \;  {\rm exp} \left [ - \frac{w}{2} \int d^4{\bf r} \;
\Delta \phi({\bf r})\, \Delta \phi({\bf r})  + 2 \,y \int d^4{\bf r} \;{\rm cos}( \phi({\bf r} ) ) 
\right ]   
\label{map}
\end{eqnarray}
\end{MyBox}
where  the two terms in the the curly bracket are summed to an exponential,  and we defined
\begin{equation}
w=\frac{1}{(2\pi)^4 \, {\cal K} } 
\label{rel1}
\end{equation}
\begin{equation}
y={\rm e}^{ - \epsilon_{_S} } 
\label{rel2}
\end{equation}

The last line of Eq.\,(\ref{map}) can be regarded as the partition function of the sine-Gordon model in $d=4$, with four derivatives and with the
parameters $w$ and $y$ related to $ {\cal K}$ and  $ \epsilon_{_S}$ of the original model by Eqs.\,(\ref{rel1},\ref{rel2}).

\section*{References}

\bibliography{Draft}

\end{document}